\documentclass[preprint,showpacs,nofootinbib,prd,aps]{revtex4-1}

\usepackage{graphicx}
\usepackage{amstext}
\usepackage{amssymb}
\usepackage[utf8]{inputenc}
\usepackage[english]{babel}
\usepackage[usenames]{color}
\usepackage[justification=centering]{caption}
\usepackage{scalerel}
\usepackage{amsmath,color}
\usepackage{dsfont}

\usepackage{stackengine,amsmath}
\stackMath
\usepackage[section]{placeins}

\begin{document}

\title{Inclusive spectra and Bose-Einstein correlations in small thermal quantum systems}
\author{M. D. Adzhymambetov$^{1}$}
\author{Yu. M. Sinyukov$^{1}$}
\affiliation{$^1$Bogolyubov Institute for Theoretical Physics,
Metrolohichna  14b, 03143 Kiev,  Ukraine}

\begin{abstract}
The spectra and correlation of identical particles emitted from small local-equilibrium sources are considered. The size of the system is defined by the negative part of the parabolic falling chemical potential. The analytical solution of the problem is found for the case of inclusive measurements. It is shown that in the case where the size of the system is comparable to the thermal wavelength of the particles, the spectra and correlation functions are far from the quasiclassical approximation expected for large systems, and observed femtoscopy scales (interferometry radii) will be essentially smaller than the Gaussian  radii of the source.  If the maximum value of the chemical potential approaches the critical one, specific for the system, one can consider the possibility of the Bose-Einstein condensation. In such a case the reduction of the intercept of the correlation function for inclusive measurements takes place. The results can be used for the searching  of femtoscopy homogeneity lengths in proton-proton collisions at  LHC energies.     
\end{abstract}

 \maketitle

\section{Introduction}
During the last few years, an intensive  femtoscopy study  of  proton-proton collisions at the LHC has been provided by the CMS \cite{CMS}, ATLAS \cite{ATLAS}, ALICE \cite{ALICE}, and LHCb \cite{LHCb} Collaborations. Some interesting results, such as the saturation of the femtoscopy scales for increasing  particle multiplicities, peculiarities of the intercept behavior for the correlation function, and  anticorrelations of identical pions were  observed. A decrease of the interferometry radii with an increase of pair  transverse momentum in $p+p$  collisions was found if a specific selection of events (e.g., according to sphericity criteria \cite{ALICE}) is not performed. One interpretation of the radii behavior is the hydrodynamization of the systems created in very high-energy $p+p$ events with large multiplicities \cite{CMS}. In this way a successive description of the femtodata on $p+p$ collisions at $\sqrt{s} = 7$ TeV in the hydrokinetic model (HKM)  has been reached in Ref. \cite{sinPLB}. This  is one of the points that allows the CMS Collaboration to interpret the obtained results \cite{CMS} for $\sqrt{s} = 13$ TeV as a consequence of hydrodynamic expansion of the thermal systems formed in $p+p$ collisions at LHC energies. 

At the same time, even if one admits  the hydrodynamic scenarios, the description of the spectra and correlations in $p+p$ collisions requires an accounting  of additional principal  aspects compared to the case of $A+A$  collisions \cite{sinPLB}. A description of the latter needs neither an uncertainty principle explicitly, nor a hypothesis about the  presence of   Bose-Einstein condensate (as for the latter, see, e.g., Refs.~\cite{Rusk,  Wong,  Flor}), nor any other “nontrivial” physics. The particle yields and their  ratios,  hadron and photon spectra, anisotropic  flows $v_n$,  quantum statistical correlation functions that bring information about the chaoticity parameter and interferometry radii,  and other observables in $A+A$ collisions are quite successfully described at the top RHIC energy and all the available LHC energies on the basis of  relativistic viscous hydrodynamics, in particular, within integrated hydrokinetic model (iHKM), see Refs. \cite{13,14,15,16,17,18,19}. The reason for the success of standard hydrodynamic and kinetic methods is that at the active stage of  spectra formation in $A+A$ collisions, the thermal/effective particle wavelengths are much smaller than the sizes of the system more precisely, than the corresponding homogeneity lengths \cite{hl1,hl2}

One of the peculiarities of the correlation femtoscopy for $p+p$ collisions is the smallness of homogeneity lengths in the strongly interacting system created in these processes: their typical effective sizes are about 1~fm, which is comparable with the mean wavelengths of emitted particles. As was considered in Ref. \cite{sinSmall}, the standard method of independent sources \cite{LLP} is violated because of the uncertainty principle: one cannot consider  the emission of the particles from different parts of a small system as independent if  the particle  wave packets  (or the regions associated  with the effective wavelengths of the quanta) are essentially overlapping.  As the result, in such cases the visible interferometry scale is reduced as compared to the geometrical system's size, and correlation function is suppressed: its intercept  decreases \cite{sinSmall}. It is worth noting, that the approach to the problem of correlation femtoscopy for small systems, developed in Ref.~\cite{sinSmall} the approach which brings a good description of the 7~TeV  $p+p$ data~\cite{sinPLB} deals, however,  with events having small and fixed multiplicity, and does not use the hypothesis of thermalization.

In this paper, we propose the results for inclusive correlation femtoscopy in an analytically solved model of small thermal quantum systems.  These findings could be applied for  correlation measurements of the homogeneity lengths \cite{hl1,hl2} in $p+p$ collisions at large mean multiplicities in a way similar to what is used in Ref. \cite{sinPLB}.

\section{Statement of the problem and basic equations}
	The main goal of the paper is to investigate the features of the inclusive spectra and correlations, which appear due to the smallness of considered quantum systems. For this purpose we apply the method of a local-equilibrium statistical operator~\cite{Zubarev}, which is a tool to obtain the density matrix $\hat{\rho}$ on the freeze-out hypersurface using the principle of maximal entropy $S(\sigma)$. Then the density matrix is defined by
\begin{equation}\label{rho0}
\hat{\rho}=\frac{1}{Z}e^{-S_{max}(\sigma)},
\end{equation}	
where $S_{max}(\sigma)$ is a maximum of entropy on the hypersurface $\sigma$ (with timelike normal vector $n_{\mu}$) under conditions fixed by the local distributions of energy, momentum, and charge density (see Ref.~\cite{2} for details). These constraints must be taken into account, for example, by the method of Lagrange multipliers. For simplicity, we consider a real free scalar field in a~$(d+1)$-dimensional space-time which is associated with the stress–energy tensor $\hat{T}^{\mu \nu}(x)=\partial^{\mu}\hat{\phi}\partial^{\nu}\hat{\phi}-\frac{1}{2}g^{\mu\nu}\left( \partial^{\rho}\hat{\phi}\partial_{\rho}\hat{\phi} - m^2 \hat{\phi}^2 \right) $ and the current of particle number density $\hat{J}^{\mu}(x)=-i\hat{\phi}^{\dagger}  \overset{\longleftrightarrow}{\partial^{\mu}} \hat{\phi}^{-}(x)$, where $\hat{\phi}^{\pm}(x)$ are the positive- and	negative-frequency parts of the field, which are defined as follows:
\begin{equation}\label{scfield}
\hat{\phi}(x)=\hat{\phi}^{\dagger}(x)+\hat{\phi}^{-}(x)=\frac{1}{(2\pi)^{d/2}}\int\frac{d^dk}{\sqrt{2k^{0}}	}		\left( a^{\dagger}_{k}e^{ikx} + a^{}_{k}e^{-ikx}\right).
\end{equation} 
Then the statistical operator [Eq. (\ref{rho0})] takes the form~\cite{2, Zubarev, Weert}:
	\begin{equation}\label{rho}
	\hat{\rho}=\frac{1}{Z}e^{ -\int d\sigma_{\nu}(x)\beta(x) n_{\mu}(x)\hat{T}^{\mu\nu}(x)+\int d\sigma_{\nu}(x)\mu(x)\beta(x)	\hat{J}^{\nu}(x) },
	\end{equation}
	where $\beta(x)=\frac{1}{T(x)}$ and $\mu(x)$ are Lagrange multipliers, corresponding to the inverse temperature and the chemical potential, respectively, and $Z$ is a corresponding partition function such that $Tr[\hat{\rho}]=1$. 	The creation and annihilation operators obey the commutation relations:
\begin{equation}\label{commutation}
[a^{}_{k_1},a^{\dagger}_{k_2}]=\delta^d(\vec{k}_1-\vec{k}_2), \qquad [a^{\dagger}_{k_1},a^{\dagger}_{k_2}]=[a^{}_{k_1},a^{}_{k_2}]=0.
\end{equation}

	Further, we consider an exact-solved model without internal flows on the hypersurface $\sigma_{\mu}$ with a uniform temperature distribution $T(x)=T$ in the moment of time $t=0$. Corresponding to the  $\sigma$ normal vector is $n_{\mu}=(1,\vec{0} )$  so $ d\sigma_{\mu}= n_{\mu}d^dx$. Thus, using Eqs. (\ref{rho}) and (\ref{scfield}), we obtain:
\begin{equation}\label{rho2}
\hat{\rho}=\frac{1}{Z}\exp \left\lbrace -\beta \int d^{d}p p^{0}a^{\dagger}_{p}a^{}_{p}+       \frac{\beta}{(2\pi)^d}\int d^{d	}x\mu(x) \frac{d^dk^{}}{\sqrt{2k^{0}}} \frac{d^dp^{}}{\sqrt{2p^{0}}} (k^0+p^{0})e^{-i(\vec{k}-\vec{p})\vec{x}} a^{\dagger}_{p} a^{}_{k}    \right\rbrace.
\end{equation}	

In the nonrelativistic limit, energy  and chemical potential can be decomposed as $p^0=m+\frac{\mathbf{p}^2}{2m}$, $\mu(x)=m+\mu_0+\mu^{'}(x)$ (restrictions for chemical potential value will be discussed later). It is easy to see that terms which contain mass $m$ in the nonrelativistic limit of Eq.~(\ref{rho2}) are reduced. For simplicity,  we take the chemical potential in parabolic form  $\mu^{'}(x)=-\sum_{i=1}^{d}\frac{{x_i}^2}{2\beta R_{i}^2}$.  

At this point, we are obliged to mention the paper \cite{Wong} which, unfortunately, we initially missed while working on the manuscript. In that article, authors consider a system of bosons in a self-consistent field of an oscillatory type. Then a “hybrid” model is constructed,  where the lowest energy level is occupied by a coherent condensate with a fixed number of particles, while the  distribution of particles over the remaining levels obeys the condition of grand canonical ensemble. Despite the similarity in mathematical formalism, such a formulation of the problem and the solution method are different  from our approach of the quasiequilibrium statistical operator corresponding to the entropy  maximum under given conditions (physical density distributions). In such a case the thermodynamic Wick theorem takes place in the system, and the chaoticity parameter is unity, which excludes a coherent condensate. The introduction of such a condensate into consideration is a specific separate problem, which we will discuss in this article later.  Further, where appropriate, we will compare the results in both approaches.

Following  Gaudin's idea~\cite{Gaudin}, modified for the case of local-equilibrium systems \cite{1}, we introduce new operators which depend on the dimensionless parameter $\alpha$ :

\begin{equation}\label{rhoalpha}
	\hat{\rho}(\alpha)=\frac{1}{Z}e^{-\alpha \beta \hat{A}}, \qquad a^{\dagger}_{k}(\alpha)=\hat{\rho}(\alpha)a^{\dagger}_{k}\hat{\rho}(\alpha)^{-1}.
\end{equation}
The operator $\hat{A}$ here is defined in the following way:
\begin{equation*}
\hat{\rho}=\frac{1}{Z}e^{-\beta \hat{A}},
\end{equation*}
\begin{equation}\label{A}
\hat{A}=\int d^{d}p \sum_{i=1}^{d}\frac{p_{i}^2}{2m}a^{\dagger}_{p}a^{}_{p}+       \frac{1}{(2\pi)^d}\int d^{d	}x\left\lbrace -\mu_0+\sum_{i=1}^{d}\frac{x_{i}^2}{2\beta R_i^2} \right\rbrace  d^dk d^dp e^{-i(\vec{k}-\vec{p})\vec{x}} a^{\dagger}_{p} a^{}_{k}
\end{equation}
Using the new operators in Eq.~(\ref{rhoalpha}), an inclusive spectrum can be calculated~\cite{Gyulassy}:
\begin{equation}\label{1inclusivespectra}
n(p)=p^{0}\frac{d^dN}{dp^d}=Tr[\hat{\rho} a^{\dagger}_{p}a^{}_{p}]=\left. Tr[a^{\dagger}_{p}(\alpha)\hat{\rho}(\alpha)a_{p}] \right|_{\alpha=1} =Tr[\hat{\rho} a^{}_{p}a^{\dagger}_{p}(\alpha=1)].
\end{equation}
	 Explicit dependence of the operator $a^{\dagger}_{p}(\alpha)$ can be obtained from the next equation, which follows from definition (\ref{rhoalpha}):
	\begin{equation}\label{eqforaalpha}
\frac{\partial a^{\dagger}_{p}(\alpha)}{\beta\partial \alpha}=[a^{\dagger}_{p}(\alpha),A].
\end{equation}	
Substituting here the expression (\ref{A}) and taking into account the commutation relations~(\ref{commutation}), we get:
	\begin{equation}\label{indifeq1}
-\frac{\partial a^{\dagger}_{k}(\alpha)}{\beta \partial \alpha}=\left( \sum_{i=1}^{d}\frac{k_{i}^2}{2m}-\mu_0\right)  a^{\dagger}_{k}(\alpha)+\frac{1}{(2\pi)^d}\int d^d x \int d^d k^{'} \sum_{i=1}^{d}\frac{x_i^2}{2 R_i^2\beta } e^{i(\vec{k}^{'}-\vec{k})\vec{x}}a^{\dagger}_{k^{'}}(\alpha).
	\end{equation}	
Here it is useful to represent the coordinates $x_i$ in the form of the derivative of the exponent with respect to momenta:
\begin{equation}
x_i^2 e^{i(\vec{k}^{'}-\vec{k})\vec{x}}=-\frac{\partial^2}{\partial k^{'2}_i}e^{i(\vec{k}^{'}-\vec{k})\vec{x}}.
\end{equation}
 Then, integrating  by parts over $k^{'}_2$ twice allows us to  integrate over $x_i$: 
\begin{equation}
-\frac{\partial a^{\dagger}_{k}(\alpha)}{\beta \partial \alpha}=\left( \sum_{i=1}^{d}\frac{k_{i}^2}{2m}-\mu_0\right)  a^{\dagger}_{k}(\alpha)   -   \frac{1}{(2\pi)^d} \int d^d k^{'}\int d^d x e^{i(\vec{k}^{'}-\vec{k})\vec{x}}  \left(\sum_{i=1}^{d}  \frac{1}{2 R_i^2\beta } \frac{\partial^2}{\partial k^{'2}_i} \right) a^{\dagger}_{k^{'}}(\alpha),
\end{equation}
	
\begin{equation}\label{finaleq}
-\frac{\partial a^{\dagger}_{k}(\alpha)}{\beta \partial \alpha}+\mu_0 a^{\dagger}_{k}(\alpha)=    \int d^d k^{'}\delta^d(\vec{k^{'}}-\vec{k}) \sum_{i=1}^{d} \left( -  \frac{1}{2 R_i^2\beta } \frac{\partial^2}{\partial k^{'2}_i} +\frac{k_i^2}{2m} \right)  a^{\dagger}_{k^{'}}(\alpha).
\end{equation}
It is our basic equation that allows uus to find solutions for inclusive thermal mean values $\left\langle a^{\dagger}_{k_1} a^{}_{k_2}\right\rangle$ that define single- and double-particle spectra in the local-equilibrium systems with a parabolic falling chemical potential.  

\section{Analytic solution of the problem}	
Since the density matrix $\hat{\rho}$~[Eq. (\ref{rho2})] acting on any state does not change its particle number, the solution of Eq. (\ref{finaleq}) can be expressed as an integral over all creation operators. Moreover, due to its linearity,  the general solution can be written as 
\begin{equation}\label{formalsolution}
a^{\dagger}_k(\alpha)=\int d^dk^{'} \sum_{n}e^{-\alpha \beta \lambda_n}C_n(\vec{k},\vec{k^{'}})a^{\dagger}_{k^{'}},
\end{equation}
where ${C_n(\vec{k},\vec{k^{'}})}$ are solutions of oscillator-like equation:
\begin{equation}\label{Cneq}
(\lambda_n+\mu_0)C_n(\vec{k},\vec{k^{'}})=\left( -  \frac{1}{2 R_i^2\beta } \frac{\partial^2}{\partial k^{2}_i} +\frac{k_i^2}{2m} \right) C_n(\vec{k},\vec{k^{'}}).
\end{equation}
Since $a^{\dagger}_k(\alpha=0)=a^{\dagger}_k$, $C_{n}(\vec{k},\vec{k^{'}})$ satisfy the additional condition:
\begin{equation}\label{Cncondition}
\sum_{n}C_{n}(\vec{k},\vec{k^{'}})=\delta^d(\vec{k}-\vec{k^{'}})
\end{equation}
From Eqs. (\ref{Cneq}) and (\ref{Cncondition}) it follows that $C_n$ can be factorized:
\begin{equation}\label{Ci}
C_n(\vec{k},\vec{k^{'}})=\sum_{\{n_i\}=0}^{\infty}\delta_{ n_1+n_2+..+n_d ,n}\prod_{i=1}^d  C_{n_i}(k_i,k^{'}_i),
\end{equation}
where $\delta_{i,j}$ is the Kronecker delta.
Besides this,  Eq.~(\ref{Cneq}) allows the separation of variables
$C_{n_i}(k_i.k^{'}_i)=A_{n_i	}(k_i^{'})f_{n_i}(k_i)$. So, in terms of the variable $k_i$, it is the Schrödinger equation for a harmonic oscillator. Its solution is represented by the Hermite functions
\begin{equation}\label{psi}
\psi_n(bx)=\frac{1}{\sqrt{2^n n!}}\left( \frac{b}{\sqrt{\pi}}\right)^{1/2}e^{-b^2x^2}H_{n}(bx), \qquad H_n(x)=(-1)^n e^{x^2}\frac{d^n}{dx^n}\left( e^{-x^2}\right),
\end{equation}
while Eq. (\ref{Cncondition}) is a completeness of the orthonormal basis 
\begin{equation}\label{ortonorm}
\sum_{n_i=0}^{\infty} \psi_{n_i}(b_ik_i)\psi_{n_i	}(b_ik_i^{'})=\delta(k_i-k_i^{'}).
\end{equation}
Then, Eqs. (\ref{Cncondition}), (\ref{Ci}), (\ref{psi}), (\ref{ortonorm}) yield
\begin{equation}
 C_{n_i}(k_i.k^{'}_i)=\psi_{n_i}(b_ik_i)\psi_{n_i	}(b_ik_i^{'}),
\end{equation}	
where $b_i^2=R_i\Lambda_T=R_i/\sqrt{mT}$, and $\Lambda_T$ is the thermal (Compton) wavelength. The index $n$ in Eq.~(\ref{Cneq}) consists of $d$ components $\left(n=\left\lbrace n_1,n_2,...,n_d \right\rbrace \right)$ running runs from $0$ to infinity, and $\lambda_n=-\mu_0+\sum_{i=1}^{d}\lambda_{n_i}$. Altogether, the following notations are used in the paper:
\begin{equation}
\lambda_{i}=\omega_i\left(n_i+\frac{1}{2}\right), \qquad \beta\omega_i=\frac{\Lambda_T}{R_i}=\frac{1}{R_i\sqrt{mT}},\qquad b_i^2=\Lambda_TR_i=\frac{R_i}{\sqrt{mT}}.
\end{equation}
Now we are ready to write the solution in Eq.~(\ref{formalsolution}) precisely: 
\begin{equation}\label{a+}
a^{\dagger}_p(\alpha)=e^{\alpha\beta\mu_0}\prod_{i=1}^{d}\left( \int d k_i \sum_{n_i=0}^{\infty} e^{-\alpha \omega_i \left( n_i+\frac{1}{2}\right) }  \psi_{n_i}(b_ip_i)\psi_{n_i}(b_ik_i)\right) a^{\dagger}_k,
\end{equation}
which allows us to calculate the inclusive spectrum [Eq.~(\ref{1inclusivespectra})]
\begin{equation}\label{2inclusivespectra}
\left\langle a^{\dagger}_{k_1}a^{}_{k_2}\right\rangle =\left\langle a^{}_{k_2} a^{\dagger}_{k_1}\left( \alpha=1\right)  \right\rangle = e^{\beta\mu_0} \prod_{i=1}^d \left(  \int dk_i M_i\left( k_{1i},k_i\right) \right)  \left\langle a_{k_2}a^{\dagger}_{k}\right\rangle.
\end{equation}
Here we introduce a kernel $M(\vec{k}_{1},\vec{k}_2)$ 
\begin{equation}\label{Mi}
M_i(k_{1i},k_i)=\sum_{n_i=0}^{\infty} e^{- \beta \omega_i \left( n_i+\frac{1}{2}\right) }  \psi_{n_i}(b_ik_{1i})\psi_{n_i}(b_ik_i), \qquad M(\vec{k}_1,\vec{k})=\prod_{i}^d M_i(k_{1i},k_i).
\end{equation}
Equation~(\ref{2inclusivespectra}),  with the commutation relations in Eq.~(\ref{commutation}) leads to the integral equation with a separable kernel with respect to the spatial components of momenta:
\begin{equation}\label{forapendix}
\left\langle a^{\dagger}_{k_1}a^{}_{k_2}\right\rangle e^{-\beta\mu_0} =   \int d^dk \prod_{i=1}^d M_i\left( k_{1i},k_i\right)   \left\langle a^{\dagger}_{k} a_{k_2}\right\rangle + \prod_{i=1}^d M_i\left( k_{1i},k_{2i}\right) .
\end{equation}
The solution of this equation can be found in the form \footnote{Do not consider $s$ as the number of particles in the system, the decomposition by which is derived in Appendix~A.}
\begin{equation}\label{3inclusive}
\left\langle a^{\dagger}_{k_1}a^{}_{k_2}\right\rangle= \sum_{s=1}^{\infty} e^{s \beta \mu_0}\prod_{i=1}^d K_i^{(s)} (k_{1i},k_{2i})=\sum_{s=1}^{\infty} e^{s \beta \mu_0}K^{(s)} (\vec{k}_{1},\vec{k}_{2})
\end{equation}
with the recurrent  equation on $ K_i^{(s)} (k_{1},k_{2})$:
\begin{equation*}
K_i^{(1)}(k_1,k_2)=M_{i}(k_1,k_2),
\end{equation*}
\begin{equation}\label{inteq}
K_i^{(s)} (k_{1i},k_{2i})  =\int dk M_i(k_{1i},k)K_i^{(s-1)}(k,k_{2i}).
\end{equation}
The kernel $M_i(k_{1},k)$ can be calculated from the definition in Eq.~(\ref{Mi}) using Mehler's formula \cite{Mehler},\cite{Magnus}:
\begin{equation}
\sum_{s=0}^{\infty}u^s\psi_s(x)\psi_s(y)=\frac{1}{\sqrt{\pi (1-u^2)}}\exp\left(  -\frac{1-u}{1+u}\frac{(x+y)^2}{4}-\frac{1+u}{1-u}\frac{(x-y)^2}{4} \right).
\end{equation}
\begin{equation}
M_i(k_{1i},k_{2i})=\sqrt{\frac{b_i^2}{2\pi \sinh(\beta \omega_i)}} \exp\left[ -(k_{1i}^2+k_{2i}^2)\frac{b_i^2\coth(\beta\omega_i)}{2}+k_{1i}k_{2i}\frac{b_i^2}{\sinh(\beta\omega_i)}\right]. 
\end{equation}
One can verify that the solution of Eq.~(\ref{inteq}) takes the form
\begin{equation}\label{Mi2}
K^{(s)}_i(k_{1i},k_{2i})=\sqrt{\frac{b_i^2}{2\pi \sinh(s\beta \omega_i)}} \exp\left[ -(k_{1i}^2+k_{2i}^2)\frac{b_i^2\coth(s\beta\omega_i)}{2}+k_{1i}k_{2i}\frac{b_i^2}{\sinh(s\beta\omega_i)}\right].
\end{equation}
Equations (\ref{3inclusive}), (\ref{inteq}), and (\ref{Mi}) lead to the expression for the inclusive spectrum, which in the variables $\vec{k}=\frac{\vec{p}_{1}+\vec{p}_{2}}{2}$ and $\vec{q}={\vec{p}_{1}-\vec{p}_{2}}$ takes the form:
\begin{equation}\label{finalspectrum}
\left\langle a^{\dagger}_{p_1} a^{}_{p_2}\right\rangle =\sum_{s=1}^{\infty}e^{\beta \mu_0 s} \prod_{i=1}^{d} \sqrt{\frac{b_i^2}{2\pi \sinh\left( s\beta \omega_i\right) }}e^{-b_i^2	k_i^2 \tanh\left(\frac{s\beta \omega_i}{2} \right) -\frac{b_i^2 q_i^2}{4}	\coth\left( \frac{s\beta \omega_i}{2}\right)}.
\end{equation}
This equation corresponds to the one  derived in Ref~\cite{Naraschewski} in the configuration representation for the trapped Bose gas. An average number of particles in the system can be obtained after integration over momentum
\begin{equation}\label{Ntot}
\left< N \right>=\int d^dp \left\langle a^{\dagger}_{p} a^{}_{p}\right\rangle=\sum_{s=1}^{\infty}e^{\beta\mu_0s} \prod_{i=1}^{d}\frac{1}{2\sinh \left( \frac{s\beta \omega_i}{2} \right)  }.
\end{equation}
A necessary condition for convergence of the series is
\begin{equation}
\lim_{s\to\infty}e^{\beta\mu_0 s} \prod_{i}\sinh(s\beta\omega_i)^{-1/2}=\lim_{s\to\infty}e^{\beta(\mu_0-\frac{d \bar{w}}{2}) s}=0,
\end{equation}
which gives a restriction for the maximum value of the chemical potential $\mu_0$: 
\begin{equation}\label{mumax}
\mu_0<\mu_{max}=\frac{d \bar{\omega}}{2}=\frac{\omega_1+\omega_2+...+\omega_d}{2}.
\end{equation}
The corresponding Wigner function can be obtained in the following way:
\begin{equation}\label{WFdef}
f_{W}(p,x)=\frac{1}{(2\pi)^d}\int d^d q \left\langle a^{\dagger}_{k+\frac{q}{2}} a^{}_{k-\frac{q}{2}} \right\rangle e^{-i \vec{q}\vec{x}},
\end{equation}
\begin{equation}\label{WF}
f_{W}(k,x)=\frac{1}{(2\pi)^d}\sum_{s=1}^{\infty}e^{\beta \mu_0 s} \prod_{i=1}^{d} \frac{1}{\cosh \left( \frac{s\beta \omega_i}{2} \right)}  \exp \left( -\left(  b_i^2	k_i^2 + x_i^2/b_i^2 \right)  \tanh\left(\frac{s\beta \omega_i}{2} \right)\right) .
\end{equation}	

This result, which follows directly from Eq.~(\ref{finalspectrum}), was earlier presented in Ref. \cite{Wong}.
One can expect that at some kind of thermodynamic limit,  when the thermal wavelength of the emitting bosons is much smaller than the homogeneity length -- source size in our case -- a quasiclassical limit for the Wigner function~(\ref{WF})  should be reached. The naive expectation is that such a function takes the form of the Bose-Einstein distribution with the corresponding coordinate-dependent chemical potential. To demonstrate this, one has to consider the thermal  (Compton) wavelengths of boson quanta $\Lambda_{T}$ to be much smaller than the size of the system, $\Lambda_{T	}/R=\beta\omega=\frac{1}{R\sqrt{mT}} \ll 1$. For simplicity we investigate the isotropic case ($R_1=\dotsb=R_d=R$). In this case, a linear approximation to the hyperbolic functions in Eq.~(\ref{WF}) can be  applied for $s$ less than some  value -- $s_0$, say -- such that $s_0\beta\omega \approx \frac{1}{2}$. Another criterion for being able to get a quasi-classical limit is non-positiveness of the chemical potential, $\mu_0 < 0$. Then, one can get from Eq.~(\ref{WF})

\begin{equation*}
f_{W}(k,x)=\frac{1}{(2\pi)^d}\left[\sum_{s=1}^{s_0}  e^{ - \left(\frac{k^2}{2mT} + \frac{x^2}{2R^2}-\frac{\mu_0}{T} \right)s }+ O\left(\frac{\Lambda_T}{R}\right) + \right. 
\end{equation*}
\begin{equation}
+\left. \sum_{s=s_0+1}^{\infty}e^{\beta \mu_0 s}  \frac{1}{\cosh^d( s \beta \omega /2)}   \exp \left( -\left(  b^2	k^2 + x^2/b^2 \right)  \tanh\left(\frac{s\beta \omega}{2} \right)\right) \right]
\end{equation}
Extending the first sum up to infinity (and subtracting  the added terms), we obtain a quasiclassical approximation with corrections that vanish in the thermodynamic limit when $\beta\mu_0 =$ const $< 0$ and $\beta \omega=\frac{\Lambda_T}{R} \to 0$:  
\begin{equation} 
f_{W,~qc}(k,x)=\frac{1}{(2\pi)^d}\sum_{s=1}^{\infty} \left(  e^{\frac{\mu_0}{T}-\frac{k^2}{2m T}-\frac{x^2}{2R^2}} \right) ^s =\frac{1}{(2\pi)^d} \frac{1}{e^{\frac{k^2}{2m T}+\frac{x^2}{2R^2}-\frac{\mu_0}{T}} - 1}
\end{equation}

\section{Femtoscopy analysis}
\subsection{Basic notations}
To investigate correlations in our model we have to calculate the  two-particle inclusive spectra \cite{Gyulassy} on the freeze-out hypersurface:
\begin{equation}\label{twoinclusive}
n(p_1,p_2)=p_1^0p_2^0\frac{d^{6}N}{dp_1^3dp_2^3}=Tr\left[ \hat{\rho} a^{\dagger}_{p_1} a^{\dagger}_{p_2} a^{}_{p_1}a^{}_{p_2} \right], \qquad n(p)= p^0\frac{d^{3}N}{dp^3}=Tr\left[ \hat{\rho} a^{\dagger}_{p} a^{}_{p} \right]
\end{equation}
\begin{equation}\label{cfdefinition}
C\left(k,q\right)=\frac{n(p_1,p_2)}{n(p_1)n(p_2)}, \qquad k=\frac{p_1+p_2}{2}, \qquad q=p_1-p_2,
\end{equation}
where  $C\left(k,q\right)$ is a correlation function (CF), which carries information about the femtoscopy scales of the system $R_{side}, R_{out}, R_{long}$. The extraction of these radii can be performed by the Gaussian fit of the CF in the low-$q$ region~\cite{Makhlin}:
\begin{equation}
C(k,q)=1+\lambda(k)e^{-R^2_{out}q^2_{out}-R^2_{side}q^2_{side}-R^2_{long}q^2_{long}}.
\end{equation}
The value of the CF at zero relative momentum $q=0$ is usually called an intercept $C(k,0)=1+\lambda(k)$, and $\lambda(k)$ is a chaoticity parameter. In numerical calculations we consider only the isotropic systems ($R_1=R_2=R_3=R$) and find the interferometry radius by fitting the one-dimensional projection of the CF (i.e. $q_1=q$, $q_2=q_3=k_2=k_3=0$) in the range of $q$ limited by the condition of $1+\lambda(k)>C(k,q)>1+0.7\lambda(k)$. The obtained interferometry radius will be addressed as $R_{HBT}$.

\subsection{Ideal Bose gas femtoscopy}
The thermal average of four operators in a noninteracting boson system in grand canonical ensemble is reduced to a sum of the products of two-operator averages by means of  Wick's theorem:
\begin{equation}\label{wicktheorem}
\left< a^{\dagger}_{p_1} a^{\dagger}_{p_2}a^{}_{p_3} a^{}_{p_4}\right> =  \left< a^{\dagger}_{p_1} a^{}_{p_3} \right>\left< a^{\dagger}_{p_2} a^{}_{p_4}\right>+\left< a^{\dagger}_{p_1} a^{}_{p_4} \right>\left< a^{\dagger}_{p_2} a^{}_{p_3}\right>
\end{equation}
Moreover, for the grand canonical ensemble of ideal Bose gas in a finite volume, the partition function of the whole ensemble factorizes over all possible energy levels which means that  Wick's theorem is applicable even for each of these levels independently.  Consequently, to examine the correlation of the system, one needs to calculate only two-operator averages: 
\begin{equation}\label{CF}
C(k,q)=\frac{\left< a^{\dagger}_{k_1}a^{}_{k_2} \right>  \left< a^{\dagger}_{k_2}a^{}_{k_1} \right> } {\left< a^{\dagger}_{k_1}a^{}_{k_1} \right> \left< a^{\dagger}_{k_2}a^{}_{k_2} \right>}+1.
\end{equation}

As a result,  in contrast to Ref.~\cite{Wong}, where the coherent condensate is postulated from the very beginning, in a pure thermal system, which is presented at the freeze-out stage as the local-equilibrium free  Bose gas, the  chaoticity parameter $\lambda(p)\equiv 1$. 
 In our approach the ground state of the  system is described by the grand canonical ensemble, which implies any number of particles occupying this state, so the consideration of a coherent condensate (if it appears)  should be different from just postulating its existence with a fixed particle number as in Ref. \cite{Wong}. We will discuss the possibility of a scenario with a coherent condensate in the next subsection.

\begin{figure}[h]
\center{\includegraphics[scale=1]{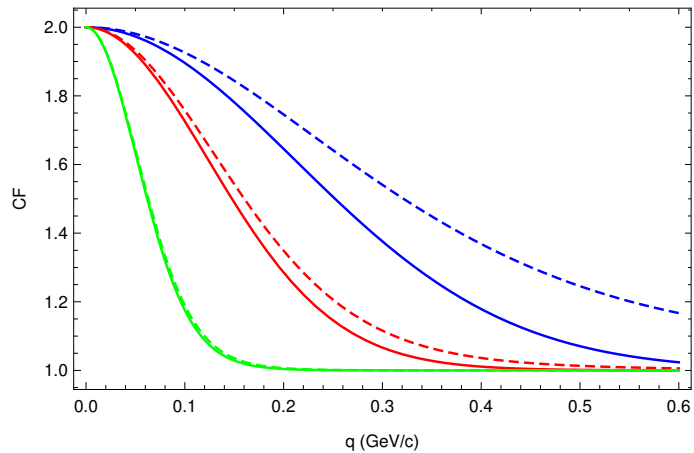}}
\caption{Quantum (solid lines) and quasiclassical (dashed lines) kaon correlation functions for  different sizes of the system at $T=155$ MeV  and $k=0.15$~GeV/$c$. The chemical potential is negative, $\mu_{0}=-0.1 \mu_{max}$ ($N\simeq 1$). Blue lines correspond to $R=0.75$~fm, red lines to $R=1.25$~fm, green lines $R=3$~fm.}
\label{fig:1}
\end{figure}

Aiming to show the  importance of quantum effects in small systems for the femtoscopy analysis,  we compare the correlation functions in quantum and quasiclassical approaches. For this purpose, the pointlike bosons with the masses of $K$ and/or  $\pi$ mesons are considered  on the  freeze-out hypersurface with the temperature $T=T_{f.o.}=155$~MeV ($\sim 10^{12}$~K); then, the thermal  wavelengths of quanta are  $\Lambda_{T}^{K}=\frac{1}{\sqrt{m_{K} T}}\approx 0.75$ fm for kaons and  $\Lambda_{T}^{\pi}\approx$1.35~fm for pions.  Figure~\ref{fig:1} shows the dependence of the correlation functions  on the relative momentum  $q$ of kaon pairs at the half-momentum $k=0.15$~GeV/$c$   and negative chemical potential $\mu_0=-0.1 \mu_{max}$  in  both approaches. Three different homogeneity scales are considered: $R=0.75$~fm, $R=1.25$~fm, and $R=3$~fm. For  sizes of about 1~fm, which are typical for  $p+p$ collisions \cite{ppALICE},  the quantum corrections are substantial as one can see. At the same time, for $R=3$~fm the corrections are fairy small, so that for the sizes typical  for  $A+A$ collisions they can be ignored, except for the case when $\mu_0 \to \mu_{max}$ [see Eq. (\ref{mumax})].

The femtoscopic analysis of  high-energy $p+p$ collisions accompanied by relatively large multiplicities  has to be carried out  more carefully, since the particle density is high ($\mu \to \mu_{max}$), and most bosons occupy the lowest energy level. In Appendix~B  the  average number of particles on this level is found and it can be described by Eq. (\ref{N0}):
\begin{equation}\label{44}
\left< N_0 \right> = \frac{1}{e^{\beta(\mu_{max}-\mu_0)}-1}.
\end{equation}
Note, that the same formula  follows generally from general Eq.~(\ref{Ntot}) when the thermal wavelengths of quanta  exceed the geometrical size of the system $\Lambda_{T}\gtrsim R$ [e.g., $T\to 0$ or $R \to 0$, after the linearization of~$\sinh (n\beta\omega)$]. The total number of bosons on this level still strongly depends on the constant part of the chemical potential $\mu_0$. It is worth noting that the momentum spectrum $\left< a^{\dagger}_{k_1}a^{}_{k_2}\right>$ of the lowest energy level [Eq.~(\ref{f0}); see Appendix~B]  factorizes over the  $k_1$ and $k_2$, so that the correlation function (\ref{CF}) in one-level approximation is constant, $C(q)=2$.  That leads to an important consequence that is shown in Fig.~\ref{fig:2}(a) -- specifically, a broadening of the complete correlation  function with the increase of the chemical potential (when the impact of the ground state increases). It means that interferometry radii obtained from the Gaussian fit of the real correlation function can be noticeably smaller  than those  formally related to the isotropic Gaussian source, $f_{G}\sim \exp(-x^2/2R^2)$, with a naive correlation function for an independent boson emission, $C(q)=1+\lambda\exp{(- R^2q^2)}$, and $\lambda=1$ for a fully chaotic emission. 

At the end of this section, let us emphasis again that even a  large number of bosons  at the lowest level in  an ideal Bose gas, concentrated in effectively limited volume and considered in the grand canonical ensemble,  does not bring coherence in the system.

\subsection{Coherence state approach}
 Now we approach a very important point:When occupation numbers at the ground state become dominant, it can lead to significant overlap between wave packets of  bosons, and  coherence in the system may develop \cite{sinSmall, Csorgo, Glauber}. The strongly overlapping bosons can hardly be considered as  fully independently emitted, and even rather small interactions between them can bring correlations of the phases of the wave packets \cite{sinSmall}.  In that case, systems with high multiplicities,  $\mu_0$ close to $\mu_{max}$, and  $ R \lesssim \Lambda_T =  \frac{1}{\sqrt{mT}}$, have to be described by a density matrix of partially coherent thermal states\footnote{ To distinguish averages with this new density matrix from those of the grand  canonical ensemble, we will use the  ${pce}$ subscript. (Averages in the grand canonical ensemble, where we think it is important, labeled with ${gce}$ subscript).  }. 

According to the general idea described in Ref.~\cite{akk_sin_ledn} (see also Ref.~\cite{Thirring})  creation (annihilation) operators at the  freeze-out stage  split into a quantum  part, associated with $q$-numbers (operators) $b_p$  and some $c$-numbers $d_p(\Delta t_f)$ where $\Delta t_f$ is the freeze-out duration time. In the case of slow adiabatic freeze-out $d_p (\Delta t_f \rightarrow \infty) \rightarrow 0$, while in the fast freeze-out scenario that takes place in $p+p$ collisions, the coherent condensate, if it appears, might give a nonzero contribution to observed spectra (see details in Ref.~\cite{akk_sin_ledn}). Since an area of our interest is small systems, it is reasonable to consider a fast freeze-out scenario  with a near simultaneous decay of the boson coherent field into free particles: $d_p(\Delta t_f \rightarrow 0)=d_p \neq 0$. 

The description of a trapped  Bose condensed gas in atomic physics is usually followed with the problem of fluctuations in the occupancy of the ground state $\bar{N}_0$ \cite{Naraschewski, Politzer}. Specifically, the description of such systems using the grand canonical ensemble predicts variance proportional to the square of this quantity: $\left< \left(N_0- \bar{N}_0\right)^2\right>_{gce}=\bar{N}_0\left(\bar{N}_0+1\right)$, which can be derived from the distribution in Eq.~(\ref{C13}). The problem appears from the discrepancy with the  experimental settings, as in the process of cooling the number of particles in the system is conserved, and  an appropriate description should be made in the canonical ensemble. Usually, the energy exchange of the system with the environment is small, and the microcanonical ensemble has to be used instead (see, for example, Ref.~\cite{Tran}).  

In the high-energy $p+p$ collisons at the LHC, we describe not a single event (single collision) with some known number of particles, but rather millions of  them in a wide range of event-by-event  multiplicities (from a few hundred).  In addition, detectors typically cannot detect the whole system formed in $p+p$ collisions, but only part of it.  In this open subsystem the energy and even net quantum numbers  fluctuate quite significantly. Moreover, the temperature at the freeze-out in these processes is about $10^{12}$ K.  Of course, this forces  us to base the inclusive measurements on the  grand canonical ensemble, possibly with some modifications. 

Instead of GCE, we propose to use a new conception of partially coherent ensemble (PCE) that is applied if the {\it mean} occupancy of the ground state exceeds some critical value $N_{c}$. The latter depends on the peculiarity of (weak) interaction in the Bose gas, that makes it not quite ideal. Depending on whether the condition $\left\langle N_0\right\rangle > N_{c}$ is satisfied, we attribute the ensemble either to the PCE or to the GCE. In the case of the PCE, the lowest state transforms into a Glauber coherent state in a way when all the mean values in the GCE and PCE are the same.\footnote{ A similar, to some extent, approach was proposed in Ref.~\cite{Wong}, where the number of bosons in the lowest energy state is {\it fixed}, while occupancies of the exited states obey GCE statistics. It that model, however, the particle number variance at the ground state and  that at $N_c$ are both zero.} 

In a partial coherent ensemble, the ground state is a Glauber coherent state, and its wave function is given  in the Fock representation by
\begin{equation}\label{cohpart}
\left|\gamma\right>  =\exp\left( -\frac{\left|\gamma \right|^2}{2}\right)\sum_{n_{\bf 0}=0}^{\infty}\frac{\gamma^{n_{\bf 0}}}{\sqrt{n_{\bf 0}!}}\left|n_{\bf 0}\right>.
\end{equation}

The description of all excited states remains the same as in the GCE. Then the action of the annihilation (creation) operator on a single ensemble element factorizes into the two parts which were discussed before:
\begin{equation}
a^{}_{p}\left| {\bf i}\right> = \left( b^{}_{p}+d^{}_p \right)\left| \gamma \right>\left| {\bf i}\right>_{ex} =  \left( b^{}_{p}+d^{}_p \right)\left| {\bf i}\right>,
\end{equation}
where the $c$-number $d_{p}$ is an eigenvalue of the annihilation operator $a^{}_{p}$ corresponding to the coherent state in Eq.~(\ref{cohpart}), and $b^{}_{p}$ quantum operator acts only on exited states. To define the $d_{p}$ number, it is more natural to use annihilation(creation) operators which decrease(increase) the occupancy numbers of a three-dimensional harmonic oscillator state $a_{\bf j}$. Such operators are connected with those in the momentum space $a^{}_{p}$ through the Hermite functions [Eq.~(\ref{psi})]:
\begin{equation}\label{cohpart2}
a^{}_{p}=\sum_{j_1,j_2,j_3=0}^{\infty}\psi_{\bf j}(p)a_{\bf j},  \qquad \psi_{\bf j}=\psi_{j_1}(b_1p_1)\psi_{j_2}(b_2p_2)\psi_{j_3}(b_3p_3).
\end{equation}
Then, according to Eqs.~(\ref{cohpart}) and (\ref{cohpart2}), the  absolute value of  $d_{p}$ can be expressed be the following average
\begin{equation}\label{gamma}
\left<\gamma\right|a^{\dagger}_{p_1}a^{}_{p_2}\left| \gamma  \right> =d^{*}_{p_1}d^{}_{p_2}= \psi_{\bf 0}(p_1)\psi_{\bf 0}(p_2) \left| \gamma \right|^2. 
\end{equation}

We, however, still did not fix the value of $\left| \gamma \right|^2$, and  to do that we postulate that the one-particle inclusive spectra [Eq.~(\ref{finalspectrum})] and the Wigner function [Eq.~(\ref{WF})] in both ensembles must be the same. This condition is satisfied if we fix $\left| \gamma \right|^2=\left<N_0\right>$ from Eq.~(\ref{N0}). When this is done, it is possible to calculate the contribution from the excitation to the inclusive spectra $\left<b^{\dagger}_{p_1}b^{}_{p_2}\right>$:
\begin{equation}\label{qp_cpnd}
\begin{matrix}
\left\langle a^{\dagger}_{p_1}a^{}_{p_2} \right\rangle_{gce}=\left\langle a^{\dagger}_{p_1}a^{}_{p_2} \right\rangle_{pce} =\left\langle \left( b^{\dagger}_{p_1}+ d_{p_1}^{*}\right) \left( b^{}_{p_2}+d^{}_{p_2} \right) \right\rangle _{pce} = d^{*}_{p_1}d^{}_{p_2} + \left\langle b^{\dagger}_{p_1}b^{}_{p_2}\right\rangle _{pce}, \\
\left< b^{\dagger}_{p_1}b^{}_{p_2}\right>_{pce}=\sum_{{\bf i}\neq {\bf 0}}\psi_{\bf i}(p_1)\psi_{\bf i}(p_2)\left<a^{\dagger}_{\bf i}a^{}_{\bf i}\right>_{pce}=\sum_{{\bf i}\neq {\bf 0}}\psi_{\bf i}(p_1)\psi_{\bf i}(p_2)\left<a^{\dagger}_{\bf i}a^{}_{\bf i}\right>_{gce}.
\end{matrix}
\end{equation}	

As we already mentioned, it is reasonable to expect that  coherence develops only if the number of bosons occupying the ground state exceeds some critical value $N_{c}$. Indeed, it is hard to imagine a coherent state of one (on average) particle.\footnote{For the coherent state described by Eq.~(\ref{cohpart}), the probability of detecting  $m$ particles obeys Poisson distribution $P(m,n)=e^{-n}\frac{n^m}{m!}$ with the average $n=\left| \gamma \right|^2$.} In this paper, we do not discuss the exact dependencies of this number on different parameters (such as the size of the system); we keep it in our numerical examples  to be fixed at $N_{c}=2$. That means that the consideration described in this subsection is  applicable, as we suggest, only when $\left< N_0 \right>>N_{c}$ [see Eq.~(\ref{N0average}) in Appendix~B]. This condition creates some restriction on the chemical potential $\mu_0$ (or  average number of particles in the whole system $\left<  N\right>$).  For example, in Figs.~\ref{fig:2}(b), \ref{fig:3}(b), and \ref{fig:4}(b), where the size ($R=1.5$~fm) and temperature  ($T=155$~MeV/$c$) are fixed, we start our description from $\left<N\right>=5$ as a minimal value which satisfies the mentioned condition. One can extract some values of $N_0$ from Table~{\ref{table:1}} using the  relation $f_0=\left<N_0\right>/\left<N\right>$ [the analytic form for $f_0$ follows from Eqs. (\ref{Ntot}) and (\ref{44})). Indeed, from the first column, it follows that $\left<N_0\right>\approx 0.4\times5=2=N_c$ particles. For the larger multiplicities, that number only grows.

\subsection{Femtoscopy in the coherent approach}

Introducing the new ensemble in the previous subsection, we break  Wick's theorem of the grand canonical ensemble, which means that  we  have to modify the correlation function defined by Eq.~(\ref{cfdefinition}). Let us rewrite the two-particle inclusive spectra Eq.~(\ref{twoinclusive}) in a representation described by Eq.~(\ref{cohpart2}):
\begin{equation}
\left\langle a^{\dagger}_{p_1} a^{\dagger}_{p_2} a^{}_{p_1}a^{}_{p_2} \right\rangle_{pce}=\sum_{\bf i,j,k,l}\psi_{\bf i}(p_1)\psi_{\bf j}(p_2)\psi_{\bf k}(p_1)\psi_{\bf l}(p_2)\left\langle a^{\dagger}_{\bf i} a^{\dagger}_{\bf j} a^{}_{\bf k}a^{}_{\bf l} \right\rangle_{pce}.
\end{equation}
 To simplify this expression, we take a few steps: We distinguish terms which involve $a^{\dagger}_{\bf 0}, a^{}_{\bf 0}$ operators, apply Wick's theorem to the other terms,\footnote{For the ideal gas each energy level can be considered as an independent grand canonical ensemble, since excited states in the introduced partially coherent and grand canonical ensemble are the same; then we can apply Wick's theorem for them, but not for the ground state.} and then express two-operator averages of excited states ($a_{\bf i}, {\bf i}\neq {\bf 0}$)  through the one-particle inclusive spectra by means of Eq. (\ref{qp_cpnd}). After  these calculations, we get
\begin{equation}
\begin{matrix}
\left\langle a^{\dagger}_{p_1} a^{\dagger}_{p_2} a^{}_{p_1}a^{}_{p_2} \right\rangle_{pce} = \left\langle a^{\dagger}_{p_1}  a^{}_{p_1} \right\rangle_{gce} \left\langle a^{\dagger}_{p_2}  a^{}_{p_2} \right\rangle_{gce} +  \left\langle a^{\dagger}_{p_1}  a^{}_{p_2} \right\rangle_{gce} \left\langle a^{\dagger}_{p_1}  a^{}_{p_2} \right\rangle_{gce} + \\
+ \left|\psi_{\bf 0}(p_1)\right|^2\left|\psi_{\bf 0}(p_2)\right|^2 \left\langle a^{\dagger}_{\bf 0} a^{\dagger}_{\bf 0} a^{}_{\bf 0}a^{}_{\bf 0} \right\rangle_{pce} -2 \left|d^{*}_{p_1}d_{p_2}\right|^2 = \\
= \left\langle a^{\dagger}_{p_1} a^{\dagger}_{p_2} a^{}_{p_1}a^{}_{p_2} \right\rangle_{gce} + \left|\psi_{\bf 0}(p_1)\right|^2\left|\psi_{\bf 0}(p_2)\right|^2 \left\langle a^{\dagger}_{\bf 0} a^{\dagger}_{\bf 0} a^{}_{\bf 0}a^{}_{\bf 0} \right\rangle_{pce} -2 \left|d^{*}_{p_1}d_{p_2}\right|^2
\end{matrix}
\end{equation}
An average of four operators on the right side of this equation is an expectation value  of $n_0^2$ taken from the coherent state [Eq.~(\ref{cohpart})]. It is known that this state is described by the Poisson distribution with both average and variance equal to  $\left| \gamma\right|^2=\left<N_0\right>$; then
\begin{equation}\label{cohcf}
\begin{matrix}
\left\langle a^{\dagger}_{\bf 0} a^{\dagger}_{\bf 0} a^{}_{\bf 0}a^{}_{\bf 0} \right\rangle_{pce}=\left<N_0\right>\left( \left<N_0\right>+1 \right) \\
\left\langle a^{\dagger}_{p_1} a^{\dagger}_{p_2} a^{}_{p_1}a^{}_{p_2} \right\rangle_{pce} =  \left\langle a^{\dagger}_{p_1} a^{\dagger}_{p_2} a^{}_{p_1}a^{}_{p_2} \right\rangle_{gce}  -\left|d^{*}_{p_1}d_{p_2}\right|^2\left(1-\frac{1}{\left<N_0\right>} \right).
\end{matrix}
\end{equation}
The last equation together with Eqs.~(\ref{twoinclusive}), (\ref{cfdefinition}), (\ref{wicktheorem}), and  (\ref{gamma}) was used in numerical calculations in  Figs.~\ref{fig:2}(b), \ref{fig:3}(b), and \ref{fig:4}(b).

\subsection{Comparison of results}
\begin{figure}[h]
\begin{center}
\includegraphics[scale=0.5]{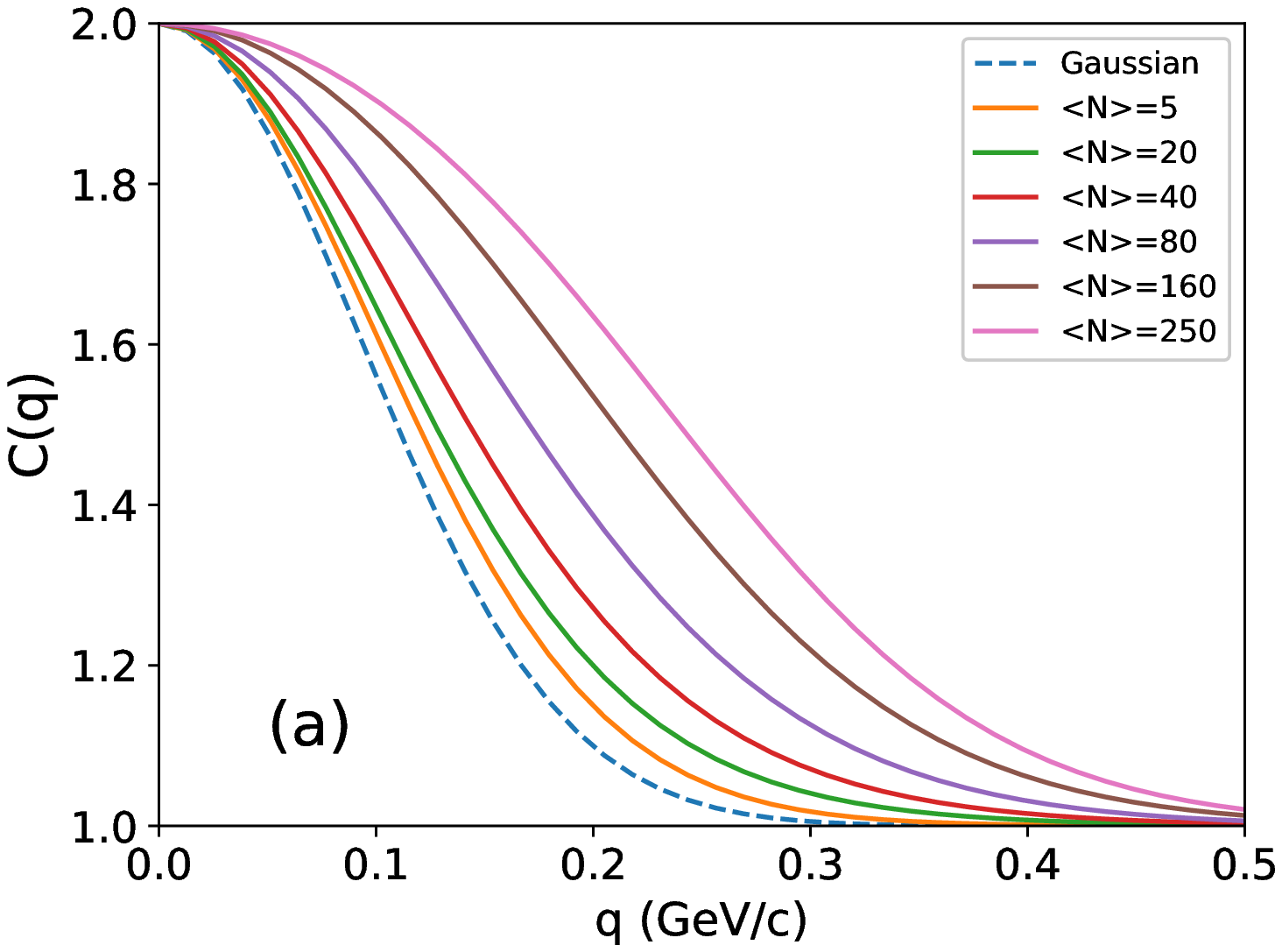}  \includegraphics[scale=0.5]{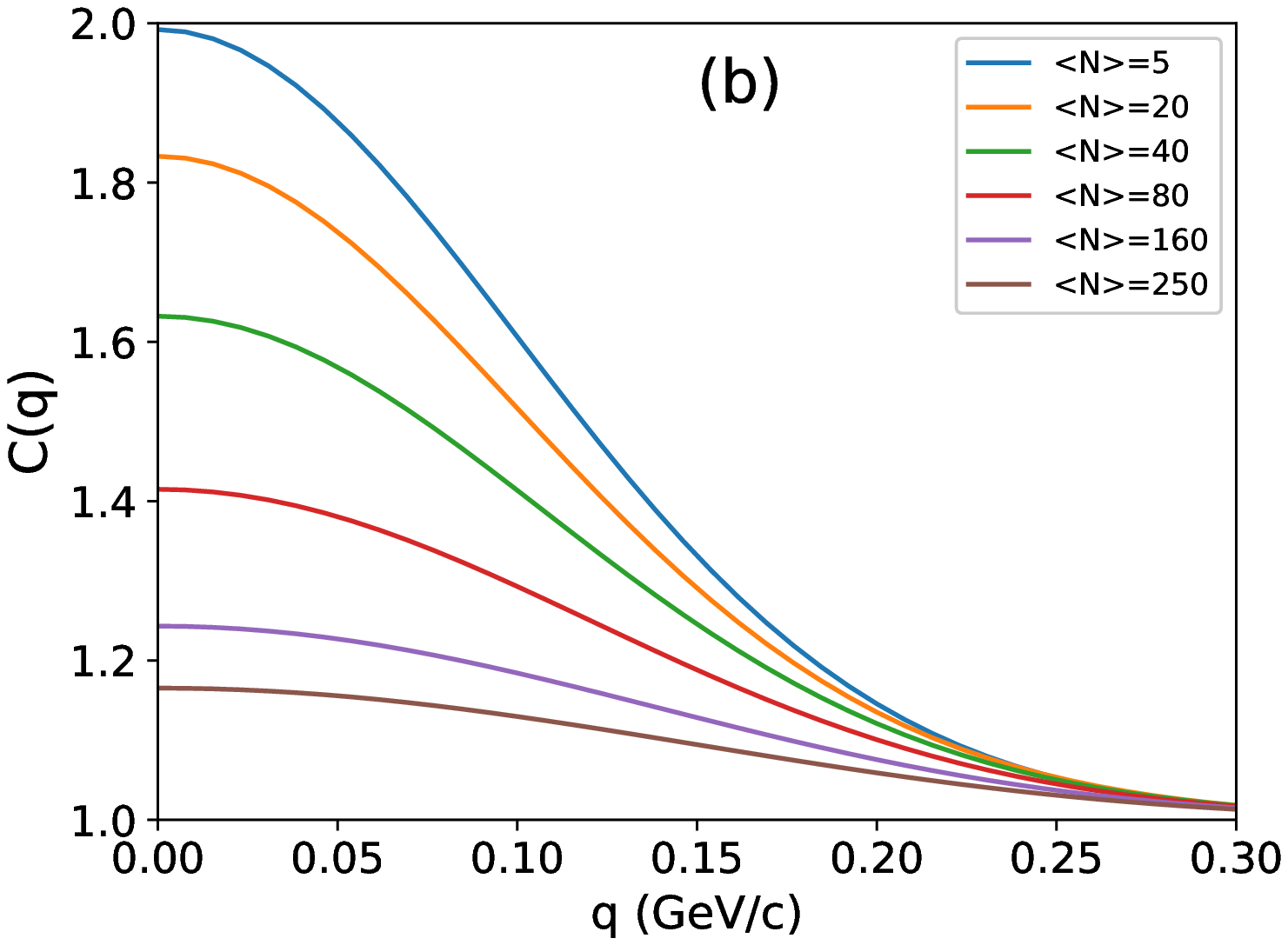} 
\end{center}
\caption{(a) Quantum statistical CFs for the different chemical potentials
with a disordered condensate (solid lines) and for a Gaussian source related to the geometrical size $R$ (blue dashed line)   at $k=0.3$~GeV/$c$, $R=1.5$~fm, and $T=155$~MeV. The solid lines correspond to  different chemical potentials, and therefore to  different average numbers of particles in the system $\left< N\right>$. (b) CF of the systems with the same $k,R,$ and $T$ as in (a) in the partial coherent state approach.}
\label{fig:2}
\end{figure}

 In Fig.~\ref{fig:2} one can see how the condensation affects the CF. For small numbers of particles, the contribution to the CF from the condensate is negligible, and the CF behaves in the same way as in chaotic systems [compare  Figs.~\ref{fig:2}(a) and~\ref{fig:2}(b)].  It is easy to see that in the case where a condensate occurs,  the intercept is less then 2 and is determined by the  condensate contribution  to the inclusive spectrum. The latter  is controlled by the constant part of the chemical potential $\mu_0$, the ratio of the thermal wavelength to the size of the system $\frac{\Lambda_T}{R}=\beta\omega$, and the average momentum of the pair $k$. 
Correlation functions in both approaches were built according the procedure described in previous subsections. Chemical potentials (see Table~\ref{table:1}) were wound numerically to guarantee proper values of $\left<N\right>$. Additionally we give corresponding condensate contributions to the spectrum $f_0=\left<N_0\right>/\left<N\right>$ which grow with the increase of multiplicity.

\begin{table}[h!]
\centering

	\begin{tabular}{|  c ||    c       |    c    |     c     | c| c| c|} 
	\hline
	$\left< N \right>$  & 5 & 20 & 40 & 80 & 160 & 250\\
	\hline
	$\frac{\mu_{max}-\mu_{0}}{\mu_{\max}}$ &    $3.02\times10^{-1}$   &    $4.69	\times10^{-2}$  & $2.09\times10^{-2}$& $9.87\times10^{-3}$ & $4.80\times10^{-3}$ & $3.04\times10^{-3}$\\ 
	\hline
	$f_{0}$ & 0.40& 0.77& 0.88& 0.94& 0.97& 0.98\\
	\hline
	\end{tabular}

\caption{Relative chemical potentials $\frac{\mu_{max}-\mu_{0}}{\mu_{\max}}$ and ground state occupancies $f_0$ at different multiplicities $\left<N\right>$ of the pion systems with $R=1.5$~fm and $T=155$~MeV. Such parameters correspond to the value $T/\omega\approx1.12$ ($\mu_{max}\approx207.6$~MeV) in Fig.~\ref{fig:3}(a).}
\label{table:1}
\end{table}

\begin{figure}[h]
\begin{center}
\includegraphics[scale=0.5]{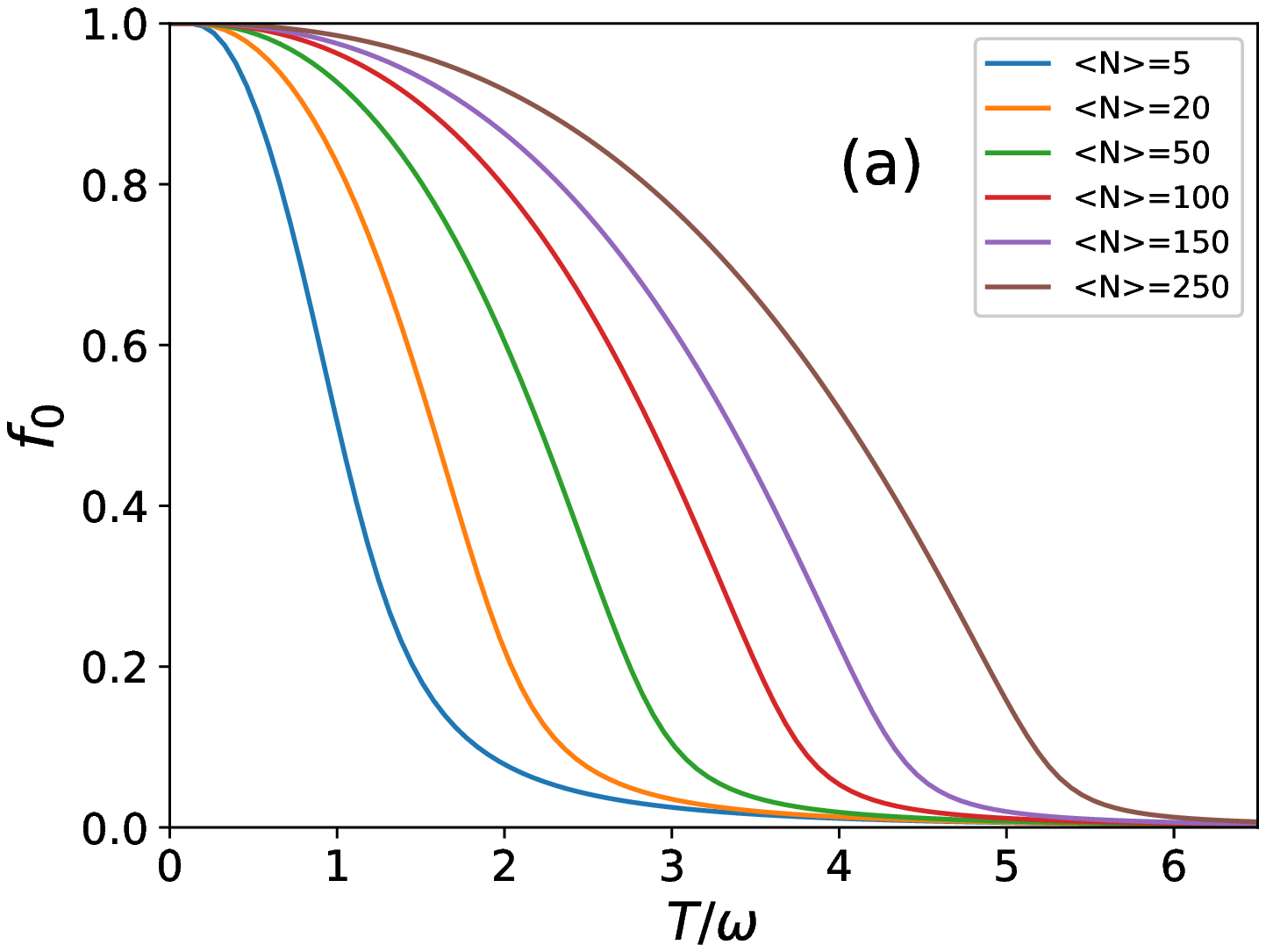}  \includegraphics[scale=0.5]{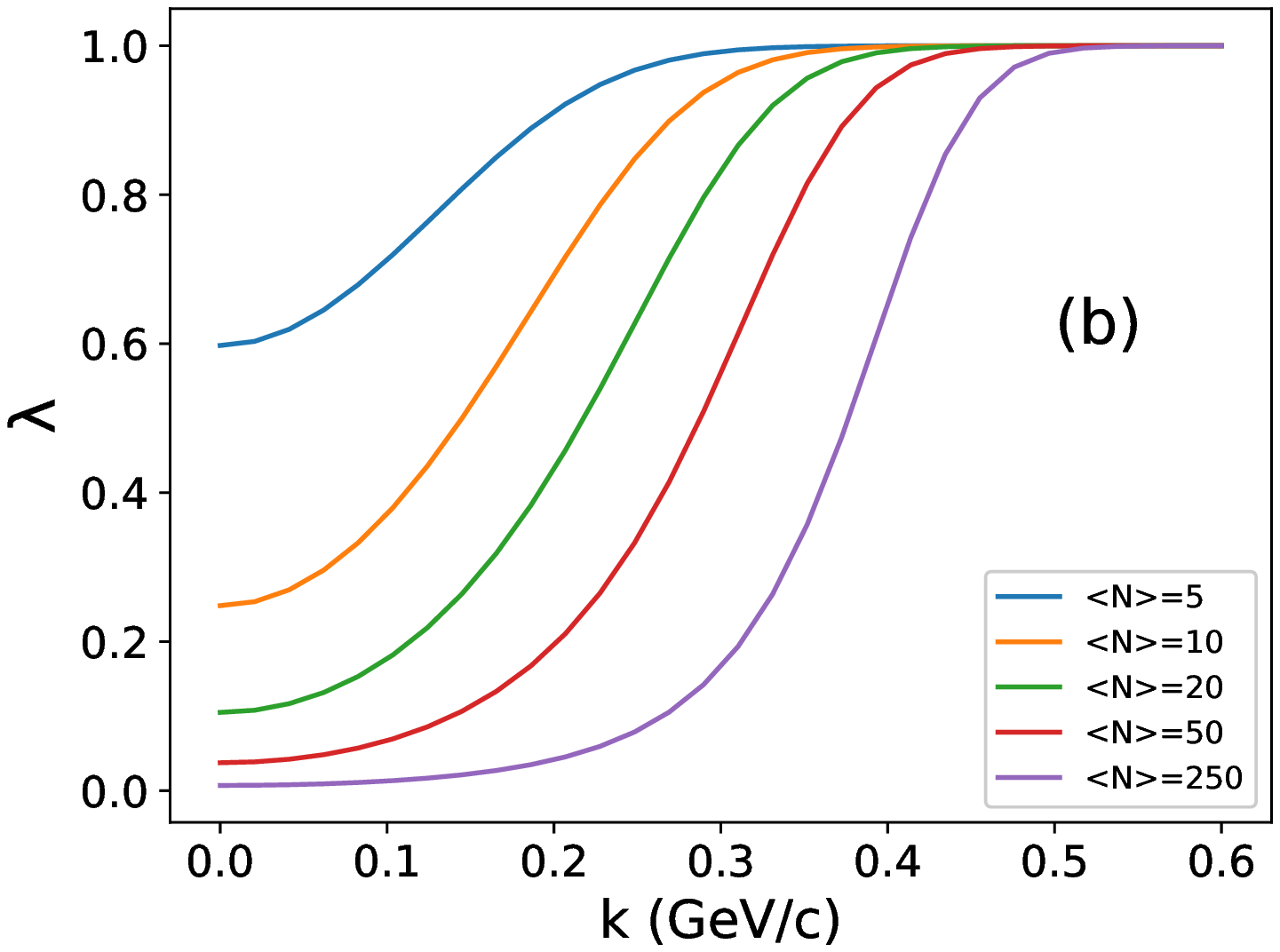} 
\end{center}
\caption{(a) The fraction of coherent condensate $f_0= \left\langle N_{0}^{coh}\right\rangle /\left\langle N \right\rangle$ as a function of $1/\beta\omega=R/\Lambda_T$ at different mean boson number $ \left\langle N \right\rangle$; (b) the $k$- dependence of chaoticity parameter $\lambda(k)$   in the grand canonical ensemble with a coherent condensate for different $\left\langle N \right\rangle$.}
\label{fig:3}
\end{figure}

In Fig.~\ref{fig:3}(b) one can see that at low $k$,  the chaoticity parameter  $\lambda(k)$ decreases in systems with a coherent condensate when $\mu_0$ approaches $\mu_{max}$. and so $\left< N \right>$  grows. It might be associated with similar experimental observations  for $p+p$ collisions  reported by the  CERN  ATLAS \cite{ATLAS} and LHCb \cite{LHCb} Collaborations. At small multiplicities the condensate contribution is small [see Fig~\ref{fig:3}(a)], and $\lambda(k)$ stays close to unity which is typical for chaotic systems. One can see from Fig.~\ref{fig:3}(b) as  was also mentioned in Ref.~\cite{Wong}, that the difference between the chaoticity parameters in  systems with high and low levels of coherence vanishes quite quickly with the increase of the momenta of measured boson pair  $k$. This happens, as follows from Eq.~(\ref{f0}),  due to the localization of  the condensate in a low kinematic region of $\sqrt{\left<k^2\right>} \sim \frac{1}{\sqrt{R\Lambda_{T}}} $, whereas the exited states shift the same average to the higher momenta. Let us mention that color lines on the plot correspond to the fixed values of $\left< N\right>$ that, however, means that the chemical potential $\mu_{0}(\left<N\right>, \beta\omega)$ has to be defined numerically for each point of the plot independently.

Figure~\ref{fig:3}(a) demonstrates  the fraction $f_0$ of the average number of particles in the coherent condensate $\left< N_{0} \right>$ compared $\left<N\right>$ for the different multiplicities available for the $p+p$ collisions at LHC, and different sizes and temperatures of the system. For the real experimental data we expect to consider sizes $R\approx 1.5$~fm, temperatures of freeze-out $T=150-165$~MeV, and multiplicities $\left< N\right> \approx 5 - 20$ identical bosons ($\pi^{\pm}$ mesons).  We, however, demonstrate much wider sets of parameters in order to compare results with $N$ in Ref. ~\cite{Wong},\footnote{In Ref.~\cite{Wong}, the ground state is occupied by the fixed number of particles.} which should coincide, since mathematically both approaches provide the same number of particles in the ground state (or the average number in our case), which certainly cannot be said about fluctuation.

 Since $T/\omega=R/\Lambda_T=R\sqrt{mT}$, the plot in Fig.~\ref{fig:3}(a) can be applied for both  $K$ and $\pi$ mesons, and one can fix $R$ to find the ``critical'' temperature $T_c(\mu)$, where $f_{0}$ becomes substantial; or fix $T$ , for example, at the typical freeze-out temperature $T_{f.o.}=155$ MeV and consider the plot as $f_0(R)$ to determine if coherence could develop in the system. 
 
\begin{figure}[h]
\begin{center}
\includegraphics[scale=0.5]{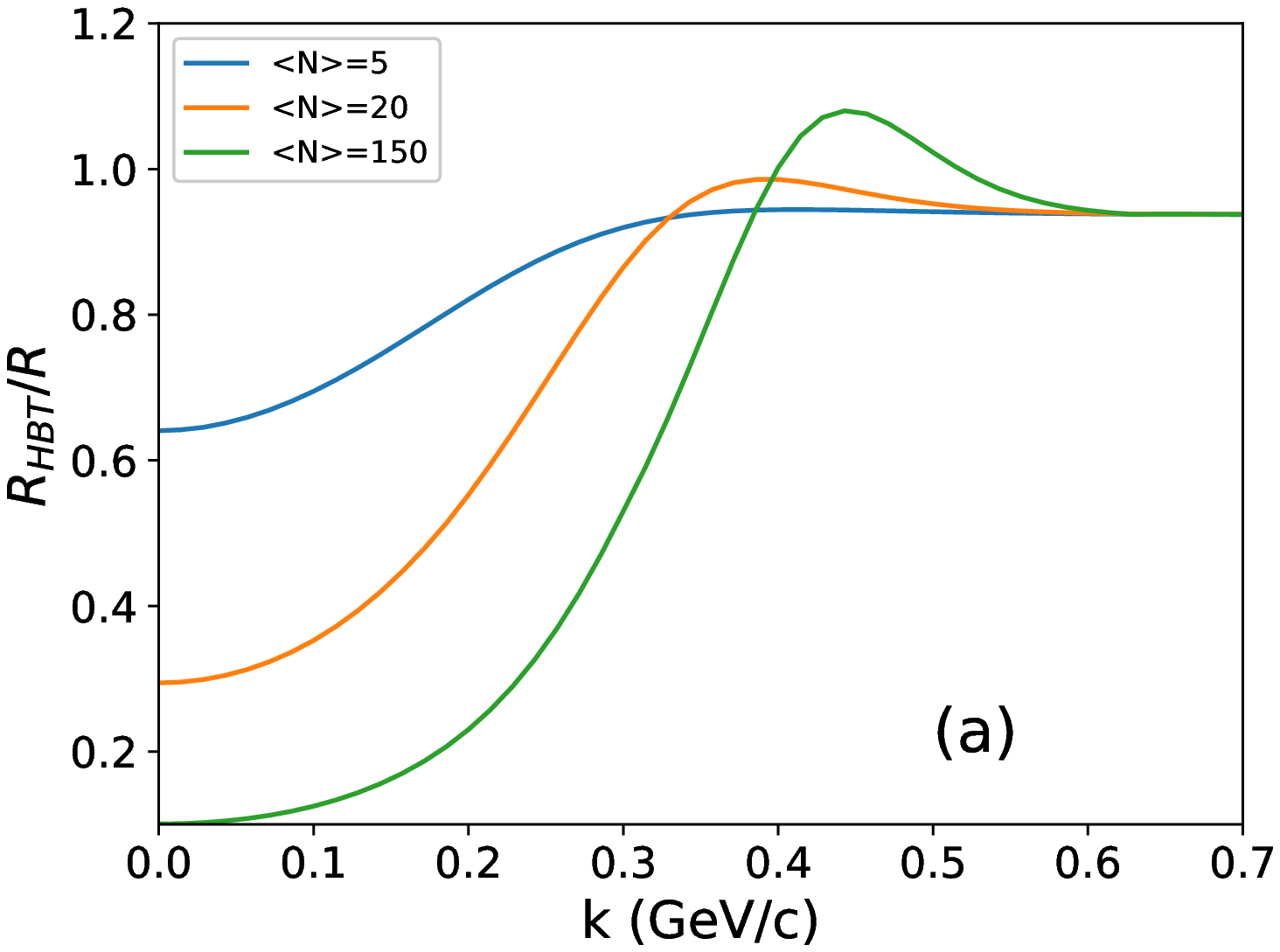}  \includegraphics[scale=0.5]{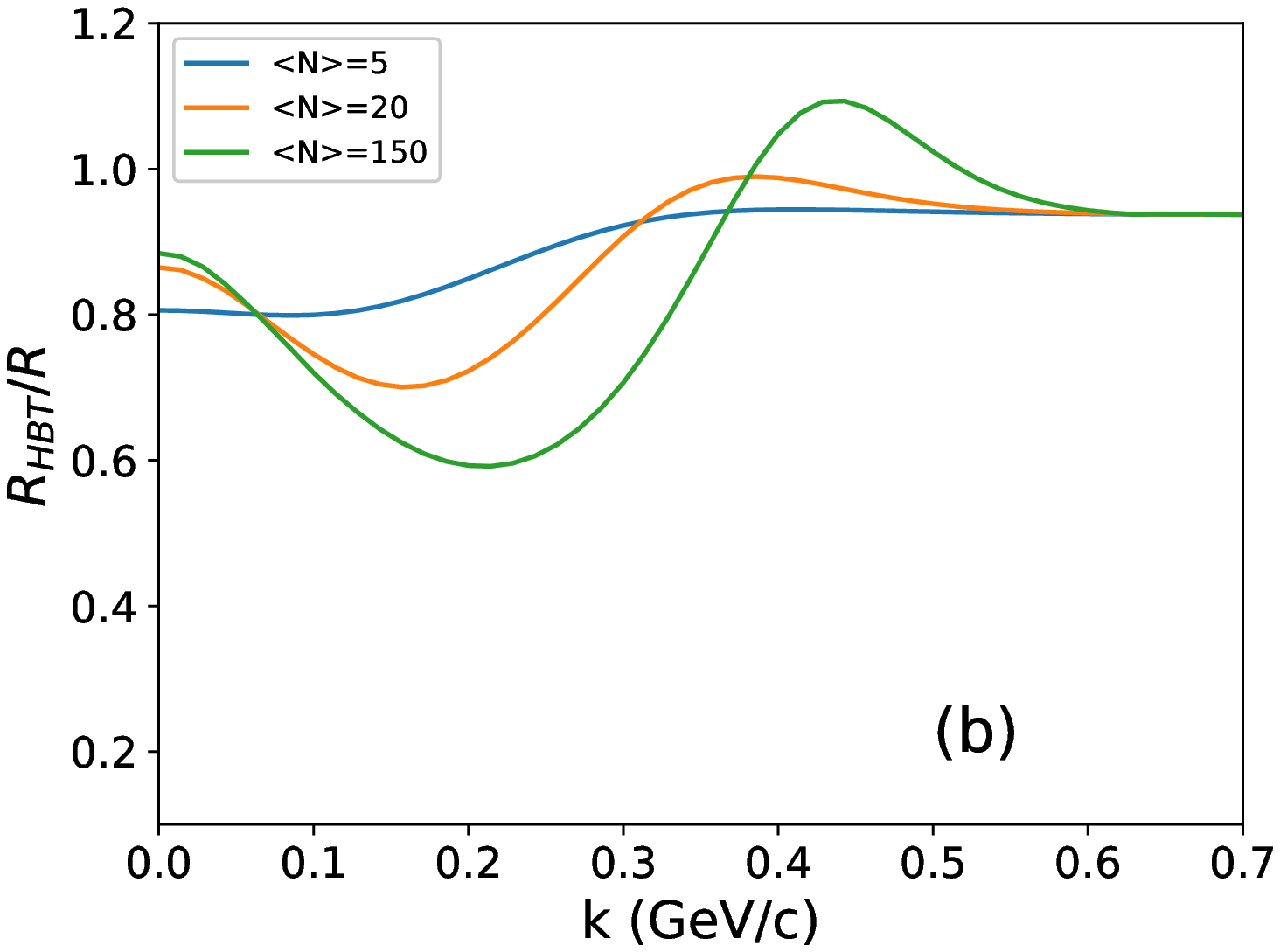} 
\end{center}
\caption{Results of the HBT fit of the pion CF at low $q$ for the small source size of $R=1.5$~fm at $T=T_{f.o.}=155 $~MeV/$c$. The plot in (a) corresponds to the fully chaotic systems, and (b) shows systems with the coherent condensate.}
\label{fig:4}
\end{figure}

 As we see in  Fig.~\ref{fig:2}, the presence of a coherent condensate changes not only intercept but also the shape of the correlation functions. The latter affects the femtoscopy radii $R_{HBT}$. In the systems with a coherent condensate, the radii $R_{HBT}$, as one can see from   Fig.~\ref{fig:4},  are  higher than in those (with the same multiplicities) where the coherence does not develop.   Also, as  is demonstrated   in Fig.~\ref{fig:4}(b), the $R_{HBT}(k)$ in the partial coherent  approach oscillate near some fixed value which can be defined from the asymptotic behavior of this dependence, while in a fully chaotic system, the femtoscopy radii at low- and high-$k$ regions can differ a lot. This happens because in low-k region, the correlation function is suppressed in the presence  of the condensate, while this is not the case in a pure ideal gas, where the  contribution from the  lowest level at small $k$ reduces the interferometry radius. At high momenta $k$, the plots in both Fig.~\ref{fig:4}(a) and \ref{fig:4}(b) to the same constant value for all multiplicities, which can be found from Eq.~(\ref{finalspectrum}) if one aborts series on the first term and neglects the condensate terms  in the CF [Eq.~(\ref{cohcf})].  Then, in this approximation
\begin{equation}
C(k,q)=C(q)=1+e^{-\frac{q^2R\Lambda_{T}}{\sinh\left( \frac{\Lambda_{T}}{R}\right)}},
\end{equation}
which corresponds to $R_{HBT}=\sqrt{\frac{R\Lambda_{T}}{\sinh\left( \frac{\Lambda_{T}}{R}\right)}}$ and $\lambda(k)=1$ [see Fig.~\ref{fig:3}(b)]. For the large systems with $R \gg \Lambda_T$, this limit can be simplified to $R_{HBT}=\sqrt{R\Lambda_{T} \left( \frac{\Lambda_{T}}{R} \right)^{-1}}=R$.

\section{Conclusions} 
 In this paper, we have studied the Bose-Einstein correlations in small local-equilibrium systems in a simple model having an exact analytic solution. We have considered a free scalar field on the freeze-out hypersurface with a uniform temperature. It is shown, that in systems, comparable in size with the thermal wavelength of emitted bosons, quantum corrections to the two-particle correlation functions of identical particles are substantial.  Qualitatively, interferometry radii of the considered systems are smaller than those formally related to the Gaussian source with the same radii as geometrical sizes of the system.  This difference increases in systems with higher multiplicities. 
 
In the case of  strong overlap of the wave packets  in the ground state in  most  of the events -- the overlapping that happens because the thermal wavelengths of the  quanta are larger or similar compared with the geometric size of the system and/or because  the chemical potential in the center of the  system approaches its maximal value  -- the coherent Bose-Einstein condensate can appear. It leads to reduction of the intercept of the inclusive correlation function. 
Note that the effect of the reduction of the femtoscales and suppression of the correlation functions compared with a naive picture of independent boson emission from a Gaussian source of the same effective size was found in a nonthermal model in Ref.~\cite{sinSmall}. Now, in the local-equilibrium thermal model, we have demonstrated in addition that  the chaoticity parameter decreases, when  the multiplicity  grows. It might be associated with similar experimental observations  for $p+p$ collisions  reported by the  CERN  ATLAS \cite{ATLAS} and LHCb \cite{LHCb} Collaborations.

The results found in this paper for the model of a small thermal source are planned to  be applied for the analysis of femtoscopic phenomena in $p+p$ collisions at the LHC.

\section{Acknowledgement}
 This research was carried out within the project ``Spatiotemporal dynamics and properties of superdense matter in relativistic collisions of nuclei, and their signatures in current experiments at the LHC, RHIC and planned FAIR, NICA". Agreement No. 7/2020 with the NAS of Ukraine. It is partially supported by the Tomsk State University Competitiveness Improvement Program.

 \appendix 

 \section{Multiplicity distribution}
One can expect that the inclusive two-point operator average (in the case of $p_1=p_2$, it is the  inclusive distribution function) $f(\vec{p}_1,\vec{p}_2)=\left< a^{\dagger}_{p_1}a^{}_{p_2} \right> $ might be expressed through the $N$-particle ``distribution functions" $f_N(\vec{p}_1,\vec{p}_2)=\left< a^{\dagger}_{p_1}a^{}_{p_2} \right>_N $. To establish this relation, we expand the grand canonical ensemble in a set of $N$-particle canonical ensembles by means of projection operators ${\cal P}^{(N)}$:
\begin{equation}
{\cal P}^{(N)}=\frac{1}{N!}\int d\vec{p}_1 \dotsb d\vec{p}_N a^{\dagger}_{p_1}\dotsb a^{\dagger}_{p_N}\left|0\right> \left< 0 \right|a_{p_1}\dotsb a_{p_N}
\end{equation}
Inserting the  completeness relation $\mathds{1}=\sum_{N=0}^{\infty}{\cal P}^{(N)}$ into Eq.~(\ref{1inclusivespectra}), we get
\begin{equation}
f(\vec{p}_1,\vec{p}_2)= \sum_{N=0}^{\infty} \left< a^{\dagger}_{p_1}a^{}_{p_2}{\cal P}^{(N)} \right> =  \sum_{N=0}^{\infty} \frac{Tr\left( \hat{\rho} {\cal P}^{(N)} \right)}{Tr\left( \hat{\rho} \right) } \frac{Tr\left( \hat{\rho} a^{\dagger}_{p_1}a^{}_{p_2} {\cal P}^{(N)} \right)}{Tr\left( \hat{\rho} {\cal P}^{(N)} \right)};
\end{equation}
note that $\frac{Tr\left( \hat{\rho} {\cal P}^{(N)} \right)}{Tr\left( \hat{\rho} \right) }=p(N) $ is a probability that the system consist of $N$ particles, and $\frac{Tr\left( \hat{\rho} a^{\dagger}_{p_1}a^{}_{p_2} {\cal P}^{(N)} \right)}{Tr\left( \hat{\rho} {\cal P}^{(N)} \right)}=f_N(\vec{p}_1,\vec{p}_2)$ is a distribution function in the canonical ensemble.  Then  
\begin{equation}\label{B3}
f(\vec{p}_1,\vec{p}_2)=\sum_{N=0}^{\infty} p(N)f_N(\vec{p}_1,\vec{p}_2). 
\end{equation}
If $p_1=p_2=p$, after integration over the momentum $p$, we get the obvious relation 
\begin{equation}
\left<N\right>=\sum_{N=0}^{\infty}p(N)N.
\end{equation}
Similarly to the calculations in the main part of the article we can derive an integral equation for $Tr\left( \hat{\rho} a^{\dagger}_{p_1}a^{}_{p_2} {\cal P}^{(N)}\right)$.  For this aim we use the following permutation relation:
\begin{equation}
{\cal P}^{(N)}a^{\dagger}_{k}=a^{\dagger}_{k}{\cal P}^{(N-1)}
\end{equation}
Strait calculations [see Eqs.~(\ref{2inclusivespectra})-(\ref{forapendix})] lead to
\begin{equation}\label{B6}
Tr\left( \hat{\rho} a^{\dagger}_{p_1}a^{}_{p_2} {\cal P}^{(N)}\right)e^{-\beta\mu_0}=\int d\vec{k} M(\vec{p}_1,\vec{k})Tr\left( \hat{\rho} a^{\dagger}_{k}a^{}_{p_2} {\cal P}^{(N-1)}\right) + Tr\left( \hat{\rho} {\cal P}^{(N-1)}\right)M(\vec{p}_1,\vec{p}_2),
\end{equation}
\begin{equation}
f_N(\vec{p}_1,\vec{p}_2)p(N)e^{-\beta\mu_0}=p(N-1) \left[ M(\vec{p}_1,\vec{p}_2)+\int d\vec{k}M(\vec{p}_1,\vec{k})f_{N-1}(\vec{k},\vec{p}_2) \right].
\end{equation}
For the case $N=1$ ($f_{0}(\vec{k}_1,\vec{k}_2)=0$), we can see that 
\begin{equation}\label{B8}
f_1(\vec{p}_1,\vec{p}_2)p(1)e^{-\beta \mu_0	}=p(0)M(\vec{p}_1,\vec{p}_2)
\end{equation}
Equations~(\ref{B8}) and (\ref{inteq}) allow us to write a solution of Eq. (\ref{B6}) in the following way:  
\begin{equation}\label{B9}
p(N)f_N(\vec{p}_1,\vec{p}_2)=\sum_{s=1}^{N}e^{s\beta\mu_0}p(N-s)K^{(s)}(\vec{p}_1,\vec{p}_2),
\end{equation}
or in terms of the number of particles in the system:
\begin{equation}\label{B10}
p(N)=\frac{1}{N}\sum_{s=1}^{N}e^{s\beta\mu_0}p(N-s)\prod_{i=1}^{d} \frac{1}{2\sinh(s\beta\omega_i/2)}, \qquad N>0,
\end{equation} 
which can be used as a recurrent equation for $p(N)$.  We can also rederive Eq.~(\ref{3inclusive}), summing up by the $N$  in Eq.~(\ref{B9}):
\begin{equation*}
\left< a^{\dagger}_{p_1}a^{}_{p_2} \right> = \sum_{N=1}^{\infty}p(N)f_N(\vec{p}_1,\vec{p}_2)= \sum_{N=1}^{\infty} \sum_{s=1}^{N}e^{s\beta\mu_0}p(N-s)K^{(s)}(\vec{p}_1,\vec{p}_2)=
\end{equation*}
\begin{equation}\label{B11}
= \sum_{s^{'}=1}^{\infty}e^{s^{'}\beta\mu_0}K^{(s^{'})}(\vec{p}_1,\vec{p}_2)\sum_{N^{'}=0}^{\infty}p(N^{'})=\sum_{s^{}=1}^{\infty}e^{s^{}\beta\mu_0}K^{(s^{})}(\vec{p}_1,\vec{p}_2).
\end{equation}

\section{Ground state spectrum}	
In order to obtain a  contribution to the boson spectrum and correlations  from the ground state, ``0'', in  the grand canonical ensemble [Eq.~(\ref{rho2})],  let us introduce the  projection operator on this state  ${\cal P}_{(0)}$
\begin{equation}
\left\langle a^{\dagger}_{k_1} a^{}_{k_2}\right\rangle_{0}= \left< a^{\dagger}_{k_1}a^{}_{k_2} {\cal P}_{(0)} \right>=\left< a^{}_{k_2} {\cal P}_{(0)} a^{\dagger}_{k_1}(\alpha=1) \right> , 
\end{equation}
\begin{equation}
  {\cal P}_{(m)}=\sum_{N=0}^{\infty}\frac{1}{N!}\left| n_1=m, \dotsb ,n_N=m \right> \left<  n_1=m, \dotsb ,n_N=m \right|
\end{equation}
The general idea in further calculations is to get an integral equation similar to Eq.~(\ref{forapendix}), but with another kernel $M_{0}(\vec{k}_{1},\vec{k}_{2})$. First, we need to determine the permutation relation between the creation operator $a^{\dagger}_{k}$ and the projection operator ${\cal P}_{(n)}$. Explicit calculations give
\begin{equation}
{\cal P}_{(n)}a^{\dagger}_{k}=\int d \vec{k}^{'} \prod_{i=1}^{d} \psi_{n}(b_i k_i) \psi_{n}(b_i k^{'}_i) a^{\dagger}_{k^{'}}{\cal P}_{(n)},
\end{equation}
\begin{equation}\label{condeq}
\left<a^{\dagger}_{k_1}a^{}_{k_2}\right>_{0}=e^{\beta \mu_0} \prod_{i=1}^{d} \left( \int dk_i M_{i,0}(k_{1i},k_i) \right)  \left( \left<a^{\dagger}_{k}a^{}_{k_2}\right>_{0} + \delta(\vec{k}_2-\vec{k}) \right),
\end{equation}
\begin{equation}
M_{i,0}(k_{1i},k_{2i})= e^{-  \frac{1}{2}\beta \omega_i }  \psi_{0}(b_ik_{1i})\psi_{0}(b_ik_{2i}), \qquad M_{0}(\vec{k}_1,\vec{k}_2)=\prod_{i}^d M_i(k_{1i},k_{2i}).
\end{equation}
Due to the orthonormality of Hermitian functions [Eq.~(\ref{psi})],  the recurrent relations [Eq.~(\ref{inteq})] simplify to the following:
\begin{equation}\label{C5}
K^{(s)}_{0}(\vec{k}_1.\vec{k}_2)=M_{0}(\vec{k}_1.\vec{k}_2),
\end{equation} 
which together with Eq. (\ref{psi}) yields
\begin{equation}\label{f0}
\left\langle a^{\dagger}_{k_1} a^{}_{k_	2}\right\rangle_{0} =\sum_{s=1}^{\infty}e^{\beta \mu_0 s} \prod_{i=1}^{d} \frac{b_i}{\sqrt{\pi} } e^{-\frac{1}{2} \beta \omega_i s} e^{-b_i^2	p_i^2  -\frac{b_i^2 q_i^2}{4}} = \frac{b_1...b_d}{\pi^{d/2}} \frac{e^{-\sum_{i=1}^{d}  b_i^2\frac{k_{1i}^2+k_{2i}^2}{2}}}{e^{\beta \left( \mu_{max}-\mu_0\right)   }-1}
\end{equation}
\begin{equation}\label{N0}
\left< N_{0} \right> = \int dp \left< a^{\dagger}_p a^{}_p \right>_{0}=\frac{1}{e^{\beta(\mu_{max}-\mu_0)}-1}
\end{equation}

As soon as macroscopical description of the lowest energy state is relevant only when it is occupied by large number of particles $N_0$, it is reasonable to find the distribution of this number in the grand canonical ensemble [Eq.~(\ref{rho2})]. For this purpose, we can use Eq.~(\ref{B10}), but with the kernels [Eq.~(\ref{C5})], which is the same as analytical continuation of the low-temperature limit,
\begin{equation}
p(N_0)=\frac{1}{N_0}\sum_{s=1}^{N_0}e^{s\beta\mu_0}p(N_0-s)\prod_{i=1}^{d} e^{-s\beta\omega_i/2}=\frac{1}{N_0}\sum_{s=1}^{N_0}e^{-s\beta(\mu_{max}-\mu_0)}p(N_0-s),
\end{equation}
which allows us to express $p(N_0)$ through $p(0)$:
\begin{equation}
p(N_0)=e^{-N_0\beta(\mu_{max}-\mu_0)}p(0).
\end{equation}
The probability $p(0)$ can be found from the normalization 
\begin{equation}
\sum_{N_0=0}^{\infty}p(N_0)=p(0)\sum_{N_0=0}^{\infty} \left( e^{-\beta(\mu_{max}-\mu_0)} \right)^{N_0}=\frac{p(0)}{1-e^{-\beta(\mu_{max}-\mu_0)}}=1,
\end{equation}
\begin{equation}\label{C13}
p(N_0)=e^{-N_0\beta(\mu_{max}-\mu_0)}\left(1-e^{-\beta(\mu_{max}-\mu_0)}\right).
\end{equation}
One can verify that 
\begin{equation}\label{N0average}
\sum_{N_0=0}^{\infty}p(N_0)N_0=\frac{1}{e^{\beta(\mu_{max}-\mu_0)}-1}=\left< N_{0} \right> 
\end{equation}
\iffalse 
Assume that the overlap between the various wave-packets  became sufficiently enough when 	the number of bosons on the same energy level exceeds some critical value $N_{c}$. The appearance frequency $P_{coh}$ of events with such property can be obtained by the next sum: 
\begin{equation}\label{C15}
P_{coh}=\sum_{N_0=N_{c}}^{\infty} p(N_0)=e^{-N_{c}\beta(\mu_{max}-\mu_0)}=\left(\frac{\left< N_0 \right>}{1+\left< N_0 \right>}\right)^{N_{c}},
\end{equation}
which approaches unity when $\left< N_0 \right> \gg 1$. We are also interested in the contribution to the inclusive spectrum only from the events with $N_0\geqslant N_{c}$, which can be obtained via Eqs. (\ref{B3}), (\ref{B9}), (\ref{C5}), and (\ref{C13}):
\begin{equation}\label{C16}
\left\langle a^{\dagger}_{k_1} a^{}_{k_2}\right\rangle_{0}^{coh} =  \left(\frac{\left< N_0 \right>}{1+\left< N_0 \right>}\right)^{N_{c}} \frac{b_1...b_d}{\pi^{d/2}} \frac{e^{-\sum_{i=1}^{d}  b_i^2\frac{k_{1i}^2+k_{2i}^2}{2}}}{e^{\beta \left( \mu_{max}-\mu_0\right)   }-1}.
\end{equation}
It has clear interpretation of the probability of appearance of  a coherent condensate in ensemble multiplied by the spectrum of the ground state (\ref{f0}):
\begin{equation}
\left\langle a^{\dagger}_{k_1} a^{}_{k_2}\right\rangle_{0}^{coh} =  P_{coh}\left\langle a^{\dagger}_{k_1} a^{}_{k_2}\right\rangle_{0}
\end{equation}
\fi

\end{document}